\documentclass[aps,prb,twocolumn]{revtex4}
\usepackage{amsfonts}
\usepackage{amsmath}
\usepackage{amssymb}
\usepackage[final,dvips]{graphicx}
\begin{document}
\title{Alternating dynamic state in
intrinsic Josephson-junction stacks self-generated  by internal resonance }
\author{A. E. Koshelev}
\affiliation{Materials Science Division, Argonne National Laboratory,
Argonne, Illinois 60439}


\begin{abstract}
Intrinsic Josephson-junction stacks realized in high-temperature
superconductors provide a very attractive base for developing coherent sources
of electromagnetic radiation in the terahertz frequency range. A promising way
to synchronize phase oscillations in all the junctions is to excite an internal
cavity resonance. We demonstrate that this resonance promotes the formation of
an alternating coherent state,  in which the system spontaneously splits into
two subsystems with different phase-oscillation patterns. There is a static
phase shift between the oscillations in the two subsystems which changes from
$0$ to $2\pi$ in a narrow region near the stack center. The oscillating
electric and magnetic fields are almost homogeneous in all the junctions. The
formation of this state promotes efficient pumping of the energy into the
cavity resonance leading to strong resonance features in the current-voltage
dependence.
\end{abstract}
\maketitle

High-temperature superconductors, such as Bi$_{2}$Sr$_{2}$CaCu$_{2}$O$_{8}$
(BSCCO), are composed of two-dimensional superconducting CuO$_2$-layers coupled
via the Josephson effect.\cite{KleinerPRL92} The large packing density of these
intrinsic junctions makes these compounds very attractive  for developing
coherent generators of electromagnetic [\textit{em}] radiation based on the
\textit{ac} Josephson effect.  Moreover, a large value of the superconducting
gap allows to bring the operation frequency of potential devices into the
practically important terahertz range. To develop a powerful source, the major
challenge is to synchronize the oscillations of the superconducting phases in a
large number of junctions. A very promising route is to excite an internal
cavity resonance in finite-size samples (mesas),\cite{LutfiSci07,KoshBulPRB08}
which can entrain oscillations in a very large number of junctions. The
frequency of this so-called in-phase Fiske mode is set by the lateral size of
the mesa. The experimental demonstration of this mechanism \cite{LutfiSci07}
has brought the quest for superconducting terahertz sources to a new level.

In general, a mechanism of pumping energy into the cavity mode is a nontrivial
issue. Homogeneous phase oscillations at zero magnetic field do not couple to
the Fiske modes.  Such coupling can be facilitated by introducing an external
modulation of the Josephson critical current density.\cite{KoshBulPRB08} In
this case the amplitudes of the generated standing wave and of the produced
radiation are proportional to the strength of modulation.

In this Letter we explore an interesting alternative possibility. Numerically
solving the  dynamics equation for the Josephson-junction stacks, we found that
near the resonance an inhomogeneous synchronized state is formed. In this
state, the system spontaneously splits into two subsystems with different
phase-oscillation patterns, formally corresponding to fluxon-antifluxon
oscillations. Inspired by numerics, we also succeeded to build such solution
analytically. The phase oscillations in two subsystems have a static phase
shift which has a soliton-shape coordinate dependence, changing from $0$ at one
side to $2\pi$ at other side. This change occurs within the narrow region near
the center of the stack and the width of this region shrinks when approaching
to the resonance. In-spite of the difference in the phase oscillation patterns
for the two subsystems, the oscillating electric and magnetic fields are almost
identical in all the junctions. The formation of this state strongly enhances
coupling to the resonance mode and promotes efficient pumping of energy into
the cavity resonance. Such state was also found recently by Lin and
Hu.\cite{LinHu08}

The dynamic equations for the Josephson-junction stacks have been derived in
Refs.\ \onlinecite{DynamEqs} and have been used in numerous simulation studies
\cite{SimulJJ}, mostly to study dynamics of the Josephson vortex lattice formed
by a magnetic field applied along the layers. We present these equations in the
form of coupled time evolution equations for reduced electric and magnetic
fields, $e_{n}$ and $h_{n}$, phase differences, $\varphi_{n}$, and the in-plane
phase gradients, $k_{n}$, which are very convenient for numerical
implementation,
\begin{subequations}
\label{DynEqsSim}
\begin{align}
\partial e_{n}/\partial\tau &  =-\nu_{c}e_{n}-g(u)\sin\varphi_{n}
+\partial h_{n}/\partial u,\\
\partial\varphi_{n}/\partial\tau &  =e_{n},\\
\partial k_{n}/\partial\tau &  =-\left( k_{n}+h_{n}-h_{n-1}\right)/\nu_{ab},\\
h_{n} &  =\ell^{2}\left(  \partial\varphi_{n}/\partial u-k_{n+1} +k_{n}\right).
%\frac{\partial e_{n}}{\partial\tau} &  =-\nu_{c}e_{n}-g(u)\sin\varphi_{n}
%+\frac{\partial h_{n}}{\partial u},\\
%\frac{\partial\varphi_{n}}{\partial\tau} &  =e_{n},\\
%\frac{\partial k_{n}}{\partial\tau} &  =-\frac{1}{\nu_{ab}}\left[  k_{n}
%+h_{n}-h_{n-1}\right],\\
%h_{n} &  =\ell^{2}\left(  \frac{\partial\varphi_{n}}{\partial u}-k_{n+1}
%+k_{n}\right).
\end{align}
\end{subequations}
The units and definitions of parameters are summarized in Table 1 and its
caption. These reduced equations depend on three parameters,
$\nu_{c}\!=\!4\pi\sigma_{c}/(\varepsilon_{c}\omega_{p})$,
$\nu_{ab}\!=\!4\pi\sigma_{ab}/(\varepsilon_{c}\omega_{p}\gamma^{2})$, and
$\ell\!=\!\lambda/s$, where $\sigma_{c}$ and $\sigma_{ab}$ are components of
the quasiparticle conductivity. We neglected the in-plane displacement current
which would give a term $\sim\partial^{2}k_{n}/\partial\tau^{2}$, because
relevant frequencies are much smaller than the in-plane plasma frequency.

We simulated a stack containing $N$ junctions ($1\!\leq \!n\!\leq N$), having a
width of $L \lambda_J$ ($0\!<\!u\!<\!L$), and assuming that the dynamic state
is homogeneous in the third direction. We study the voltage range corresponding
to the Josephson frequencies close to the lowest in-phase resonance frequency
$\omega_{1}=\pi \ell/L$. The function $g(u)\!=\!1\!-\!2r(u\!-\!L/2)/L$ in Eq.\
(\ref{DynEqsSim}a) describes a linear modulation of the Josephson current
density, which provides coupling to this mode for c-axis homogeneous
oscillations.\cite{KoshBulPRB08} Our purpose is to probe the qualitative
structure of the resonance states and, for simplicity, we assume nonradiative
boundary conditions at the edges,  $k_{n} =0$, $\partial\varphi_{n}/\partial u
= \mp I/2\ell^{2}$, for $u=0,L$, where $I$ is the transport current flowing
through the stack, and metallic contacts at the top and the bottom,
$k_{0}=k_{N+1}=0$.

\begin{widetext}
\noindent\textbf{Table 1.\ }\emph{Units of physical variables.} Here
$\omega_{p}$ is plasma frequency, $\lambda_{J}$ is the Josephson length, $s$ is
the interlayer period, $\lambda$ is the in-plane London penetration depth,
$\gamma$ is the anisotropy factor, and $j_{J}$ is the Josephson current
density.
\begin{tabular}
[t]{|c|c|c|c|c|c|c|}\hline \textbf{Variable} & time, $\tau$ & coordinate, $u$ &
phase gradient, $k_{n}$ & electric field, $e_{n}$ & magnetic field, $h_{n}$&
current density, $j$
\\\hline \textbf{Unit} & $1/\omega_{p}$ & $\lambda_{J}$ &
$1/\lambda_{J}$ & $\Phi _{0}\omega_{p}/(2\pi cs)$ &
$\Phi_{0}/(2\pi\gamma\lambda^{2})$& $j_{J}$\\\hline
\end{tabular}
\smallskip
\end{widetext}

\begin{figure}[ptb]
\begin{center}
\includegraphics[width=3.4in]{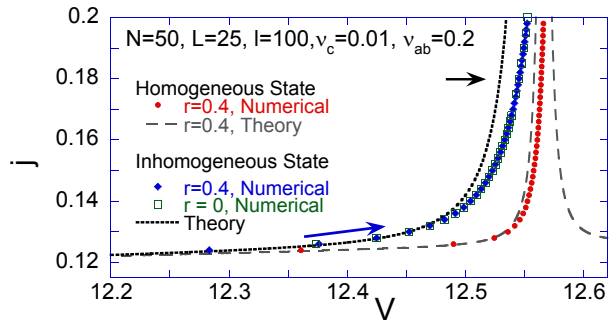}
\end{center}
\caption{Simulated and theoretical CVDs in the vicinity of the resonance
voltage for different states with $V\!=\!\langle e_n \rangle\!=\!\omega$. Small
circles show the dependence obtained for $r=0.4$ and c-axis homogeneous initial
state, which remains homogeneous with increasing current. The theoretical CVD
for this state\cite{KoshBulPRB08} is shown by grey dashed line. Small filled
diamonds show the CVD for $r=0.4$ and inhomogeneous state, which was
self-generated when small $n$-dependent perturbation was added to the phase in
the beginning of run for each current value. The open squares represents the
CVD for $r\!=\!0$, when the inhomogeneous state has been used as the initial
state at starting current. The inhomogeneous state is not sensitive to
modulation. Dotted black line shows the theoretical curve for the inhomogeneous
state based on Eqs.\ (\ref{ResCurr}) and (\ref{CouplCnstRes}).}
\label{Fig-SimulIV}
\end{figure}
\begin{figure}[ptb]
\begin{center}
\includegraphics[width=3.3in]{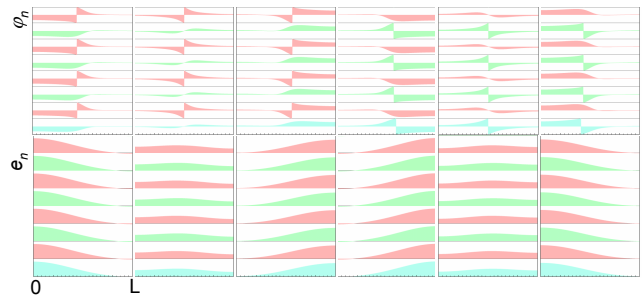}
\end{center}
\caption{Snapshots of the phase and electric field configurations in the 8
bottom junctions for the same parameters as in Fig.\ \ref{Fig-SimulIV} and
$j=0.18$ marked by the arrow.\cite{Animations} The phases are reduced to the
interval $[-\pi, \pi]$, so that jumps formally correspond to centers of the
Josephson vortices. The dynamic state corresponds to alternating nucleation,
motion and annihilation of vortices in the even junctions and antivortices in
the odd junctions.  Vortex velocities near the edges much larger than near the
center. In between $3^{\mathrm rd}$ and $4^{\mathrm th}$ configurations
jumplike annihilation of fluxons and nucleation of antifluxons take place at
the right side. The lower plots show that the electric field is homogeneous in
all junctions and has space/time dependence corresponding to the fundamental
mode. The annihilation/nucleation events correspond to maxima of electric
fields at the edges. } \label{Fig-TimeConfigs}
\end{figure}
\begin{figure}[ptb]
\begin{center}
\includegraphics[width=2.8in]{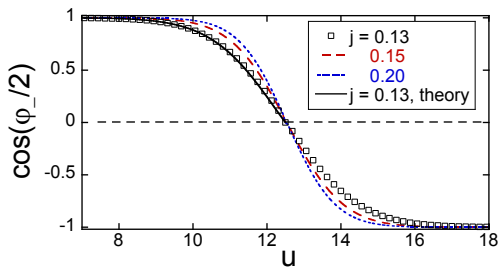}
\end{center}
\caption{Cosine of half static phase shift between phases in two subsystems at
different currents. One can see that the phase shift has shape of soliton and
it narrows with approaching to the resonance. Solid line shows theoretical
curve based on Eq.\ (\ref{Interp-phim}) and using Eqs.\ (\ref{C1}), (\ref{ls}),
and (\ref{CouplCnstRes}) for $j=0.13$. This function determines the coupling of
the homogeneous Josephson oscillations to the internal resonance mode.}
\label{Fig-Cos-df}
\end{figure}

Figure \ref{Fig-SimulIV} shows the current-voltage dependences (CVDs) obtained
for representative system parameters listed in the plot and for two values of
the modulation parameter $r=0$ and $0.4$. The dependences have been obtained
with increasing current. We observe a strong resonance enhancement of the
current due to the excitation of the internal cavity resonance. As our
simulations do not take into account thermal noise, the emerging state is
sensitive to the initial configuration. If we start with a c-axis homogeneous
state it remains homogeneous up to certain current. In this case the resonance
is excited due to the finite modulation and it is well described by theory
developed in Ref.\ \onlinecite{KoshBulPRB08}. However, if we add a small
$n$-dependent perturbation to the phase at the start of every current run, the
homogeneous state blows up and the system organizes itself into a coherent
inhomogeneous state. We also studied a system without modulation using an
inhomogeneous state as initial state and found that the corresponding CVD is
practically undistinguishable from the one for the modulated system. Therefore,
the modulation of the critical current density triggers the transition to the
inhomogeneous state but once being formed, this state is not sensitive to the
modulation any more.

To understand the nature of the inhomogeneous state, we show in Fig.\
\ref{Fig-TimeConfigs} the time evolution of the phase and electric field for
the 8 bottom junctions for $j=0.18$. We see that the system splits into two
alternating subsystems with different phase dynamics, corresponding to
fluxon/antifluxon oscillations. In the first half period vortices nucleate at
the left side in even junctions, rapidly move to the center, then, after slow
motion near the center, rapidly annihilate at the right side. Immediately after
that, in the second half period, antivortices nucleate at the right side in the
odd junctions, move to the left in a similar way, and annihilate at the left
side. In spite of the difference in the phase dynamics between the two
subsystems, the electric and magnetic fields are almost identical in all
junctions. For the electric field, this can be seen from the lower plots of
Fig.\ \ref{Fig-TimeConfigs}. The dominating contribution to the oscillating
electric field is given by the fundamental cavity mode, $e_{n}\propto \cos(\pi
u/L)$.

The homogeneity of the electric field implies that there is a static phase
shift between the phase oscillations in the two subsystems. Figure
\ref{Fig-Cos-df} shows the cosine of half of this phase shift at different
currents. One can see that the phase shift has the shape of a soliton and its
width shrinks with increasing current.

To study the dynamic state analytically, we assume that the system is split
into two alternating subsystems, $\varphi_{2m+1}=\varphi_{1}$,
$\varphi_{2m}=\varphi_{2}$.\cite{footnote1122} Introducing new variables,
$\varphi_{+}=(\varphi_{1}+\varphi_{2})/2$ and $\varphi_{-}=\varphi
_{2}-\varphi_{1}$, and excluding other variables, we derive from Eqs.\
(\ref{DynEqsSim}) for $\ell \gg 1$
\begin{subequations}
\begin{align}
\frac{\partial^{2}\varphi_{+}}{\partial\tau^{2}}&+\nu_{c}\frac{\partial
\varphi_{+}}{\partial\tau}-\ell^{2}\frac{\partial^{2}\varphi_{+}}{\partial
u^{2}}    =-\sin\varphi_{+}\cos\left(  \varphi_{-}/2\right), \label{eq-phip}\\
\frac{\partial^{2}\varphi_{-}}{\partial\tau^{2}}&+\nu_{c}\frac{\partial
\varphi_{-}}{\partial\tau}-\frac{1}{4}\left(  1+\nu_{ab}\frac{\partial
}{\partial\tau}\right)  \frac{\partial^{2}\varphi_{-}}{\partial u^{2}}
\nonumber\\
&=-2\sin\left(  \varphi_{-}/2\right)  \cos\varphi_{+}. \label{eq-phim}
\end{align}
\end{subequations}
We now obtain a self-consistent approximate solution of these equations for the
dynamic state when the Josephson frequency, $\omega =\langle e_n \rangle$, is
close to the resonance frequency, $\omega_{1}=\pi\ell/L$. We will show that
$\varphi_{-}$ is almost static. In this case the equation for $\varphi_{+}$
coincides with the phase equation for the Josephson junction with modulated
Josephson current density \cite{KoshBulPRB08} with modulation function
$g(u)\!=\!\cos\left( \varphi_{-}/2\right) $, see Fig.\ \ref{Fig-Cos-df}. Near
the resonance frequency, we make the mode projection for $\varphi_{+}$,
\begin{equation}
\varphi_{+}(u,\tau)=\omega\tau+\operatorname{Re}[\psi\exp(-i\omega\tau
)]\cos(\pi u/L), \label{ModeProj}
\end{equation}
and, assuming $|\psi|\ll1$, we obtain
\begin{align}
\psi&=\frac{ig_{-}}{\omega^{2}-\omega_{1}^{2}+i\nu_{c}\omega},
\label{ModeAmpl}\\
\text{with  }g_{-}&=\frac{2}{L}\int_{0}^{L}\cos(\pi u/L)\cos\left(
\varphi_{-}/2\right) du \label{CouplCnst}
\end{align}
being the coupling parameter. This solution determines the CVD, which takes
into account resonance enhancement of the Josephson current\cite{KoshBulPRB08},
%\delta j\approx \left\langle \cos\left(  \varphi_{-}/2\right)\sin\varphi _{+}\right\rangle
\begin{equation}
j(V)\approx\nu_{c}V+ \frac{g_{-}^{2}\nu_{c}V/4}{\left( \omega
_{1}^{2}-V^{2}\right)  ^{2}+\left(  \nu_{c}V\right)  ^{2}}. \label{ResCurr}
\end{equation}

To evaluate $\varphi_{-}(u,\tau)$, we split it into static and dynamic parts,
$\varphi_{-}(u,\tau)=\bar{\varphi}_{-}(u)+\tilde{\varphi}_{-}(u,\tau)$. Further
analysis shows that $\tilde{\varphi}_{-}(u,\tau)\ll1$. The static part is
determined by
\begin{equation}
%\frac{d^{2}\varphi_{-}}{du^{2}}-8C_{+}(u)\sin \frac{\varphi_{-}}{2}  =0
d^{2}\varphi_{-}/du^{2}-8C_{+}(u)\sin (\varphi_{-}/2)  =0 \label{Eq-phim-stat}
\end{equation}
with $C_{+}(u)\!\equiv\!\left\langle \cos\varphi_{+}\right\rangle $. Using Eqs.
(\ref{ModeProj}) and (\ref{ModeAmpl}), we obtain
\begin{align}
C_{+}(u)  &  \approx C_{1}\cos(\pi u/L),\\
\text{with }C_{1}  &  =-\frac{\operatorname{Im}[\psi]}{2}=-\frac{g_{-}}
{2}\frac{\omega^{2}-\omega_{1}^{2}}{\left(  \omega^{2}-\omega_{1}^{2}\right)
^{2}+\left(  \nu_{c}\omega\right)  ^{2}}. \label{C1}
\end{align}
Consider the region near the midpoint, $u=L/2$, where the static cosine can be
approximated by the linear function $C_{+}(u)\approx C_{1}(\pi/L)(L/2-u)$.
Using the substitution %$v=(u-L/2)/l_{s}$ with
\begin{equation}
v=(u-L/2)/l_{s} \text{ with } l_{s}=\left[  L/\left(  8\pi C_{1}\right)
\right] ^{1/3}, \label{ls}
\end{equation}
we can reduce Eq.\ (\ref{Eq-phim-stat}) near the midpoint to the dimensionless
form
\begin{equation}
d^{2}\varphi_{-}/dv^{2}+v\sin(\varphi_{-}/2)  =0. \label{Phim-Red}
\end{equation}
This equation allows for soliton solution in which $\varphi_{-}$ changes from
$0$ to $2\pi$ within $|v|\sim1$, corresponding to $|u-L/2|\sim l_{s}$. In the
case $l_{s}\ll L$ \ the linear expansion for $C_{+}(u)$ is valid in the soliton
core and Eq.\ (\ref{Phim-Red}) determines its shape with a very good accuracy.
In the range $\omega_{1}^{2}\!-\!\omega^{2}\gg\nu_{c}\omega$ the condition
$l_{s}\!\ll\! L$ is equivalent to $\omega_{1}^{4}(1\!-\!(\omega/\omega
_{1})^{2})\!\ll\!4\pi^{2}\ell^{2}$. As typically $\ell\approx 150$, this
condition is always satisfied in the interesting frequency range. The soliton
solution has the symmetry $\varphi_{-}(v)=2\pi-\varphi_{-}(-v)$. Numerically
solving Eq.\ (\ref{Phim-Red}), we can interpolate the solution as
\begin{equation}
\varphi_{-}(v)\approx \pi\exp \left\{  -(\sqrt{2}/3)
\left[\left(|v|\!+\!C_v\right)^{3/2}\!-C_v^{3/2}\right] \right\}
\label{Interp-phim}
\end{equation}
for $v\!<\!0$ with $C_v\approx 0.5129$.

To evaluate the coupling constant (\ref{CouplCnst}), we note that $\cos\left(
\varphi_{-}/2\right)$ changes from $1$ to $-1$ within a narrow region near the
midpoint, $|u\!-\!L/2|\!\sim\! l_{s}$ meaning that, up to terms $\sim\!\left(
l_{s}/L\right)^{2}$, $\cos(\varphi_{-}/2)$ can be approximated by
$\mathrm{sign}(L/2\!-\!u)$ which gives $g_{-}\!\approx\! 4/\pi$. Surprisingly,
this self-generated steplike modulation provides the maximum possible coupling
to the resonance mode. Correction to $g_{-}$
%of the order of $\left( l_{s}/L\right) ^{2}$
due to the finite soliton width can be evaluated as
\[
\delta g_{-}\!=\!\frac{4}{L}\int_{0}^{L/2}\!\cos\!\left(\frac{\pi
u}{L}\right)\left(\cos\! \frac{\varphi_{-}}{2}-\!1\right)
\!du\!\approx\!-0.464\frac{4\pi l_{s}^{2}}{L^{2}}.
\]
Adding this correction and using Eqs.\ (\ref{ls}) and (\ref{C1}), the total
coupling parameter can be written as
\begin{equation}
g_{-}\approx\frac{4}{\pi}\left\{  1-1.817\left[  \frac{\left(  \omega_{1}^{2}
-\omega^{2}\right)  L^{2}}{\left(  \omega^{2}-\omega_{1}^{2}\right) ^{2}+\left(
\nu_{c}\omega\right)  ^{2}}\right]  ^{-2/3}\right\} \label{CouplCnstRes}
\end{equation}
As $g_{-}\sim 1$, the used linear approximation $|\psi|\ll 1$ is valid only at
$|\omega^2-\omega_{1}^2|>1$.

To evaluate the time-dependent part of $\varphi_{-}$, we represent it in the
complex form, $\tilde{\varphi }_{-}(\tau,u)=\operatorname{Re}\left[
\tilde{\varphi}_{-}(u)\exp(-i\omega \tau)\right]  $, and separating the
time-dependent part of Eq.\ (\ref{eq-phim}),  we derive the equation for the
complex amplitude
\[
\left(  \omega^{2}+i\nu_{c}\omega\right)
\tilde{\varphi}_{-}+\frac{1\!-\!i\nu_{ab}\omega} {4}
\frac{\partial^{2}\tilde{\varphi}_{-} }{\partial
u^{2}}=2\sin\frac{\bar{\varphi}_{-}}{2}
\]
In the range $\omega\!\gg\!1$, we estimate $
\tilde{\varphi}_{-}\!\approx\!2\sin\left(
\bar{\varphi}_{-}/2\right)/\omega^{2}\!\ll\!1 $. The small amplitude of
$\tilde{\varphi}_{-}(\tau,u)$ justifies usage of the static approximation for
$\varphi_{-}$ in the equation for $\varphi_{+}$.

To compare analytical results with numerics, we show in Fig.\ \ref{Fig-Cos-df}
the theoretical result for $\cos(\varphi_{-}/2)$ at $j\!=\!0.13$ based on Eq.\
(\ref{Interp-phim}). It accurately describes the numerical data. The
theoretical prediction for the CVD based on Eqs.\ (\ref{ResCurr}) and
(\ref{CouplCnstRes}) is shown in Fig.\ \ref{Fig-SimulIV}. The linear
approximation describes well the numerical data for voltages not too close to
the resonance. Due to the enhancement of nonlinearities in the vicinity of the
resonance, the analytical result overestimates the current increase.

The found state looks similar to the fluxon-antifluxon oscillations in a single
junction.\cite{FultonDynesSSC73} These oscillations appear as a result of a
parametric instability of the homogeneous oscillations \cite{CostabilePRB87}
and lead to so-called zero-field steps in the CVDs.\cite{Chen71} The Josephson
frequency of such step is twice the frequency of the involved resonance. In
spite of the apparent similarity, there are essential qualitative differences.
In the case of a stack, the frequency coincides with the resonance frequency.
The dynamic configurations are also very different. In the case of a single
junction, a well-developed fluxon nucleates at one side, moves with Swihart
velocity to the other side, converts to the antifluxon there, which then moves
back again with constant velocity \cite{FultonDynesSSC73}. In our case, there
is a region statically located near the center where rapid phase change $\pm
\pi$ is localized, corresponding to $2\pi$ phase change of $\phi_{-}$.  As a
consequence, the centers of fluxons and antifluxons, formally defined as points
where the phases are commensurate with $\pm \pi$, spend most time near the
center and very rapidly jump to and from the edges, see Fig.\
\ref{Fig-TimeConfigs}. Moreover, the fluxon interpretation of our oscillations
is somewhat artificial, as there are no well-defined localized soliton
excitations moving across the junctions.

The alternating state is a plausible candidate for the coherent state
responsible for resonant terahertz emission observed in Ref.\
\onlinecite{LutfiSci07}. Even though we did not use boundary conditions
accounting for the radiation, it is clear that the generation of such a state
would lead to powerful emission. In fact, the radiation does not influence much
the structure of the internal states for short mesas, $N\lesssim 100$, and
\textit{em} emission can be approximately computed from the oscillating
electric fields at the edges.\cite{KoshBulPRB08} For taller mesas, the
radiation may contribute to the resonance damping but we do not expect that it
will destroy the coherent state. The experimental resonance features in the
CVDs are much weaker then the theoretical ones. The possible mechanisms
reducing the amplitude of the resonance include noise, c-axis inhomogeneities,
and additional damping channels not taken into account by the theoretical
model.

The author would like to thank U.\ Welp, L.\ Bulaevskii, K.\ Gray, M.\ Tachiki,
and X.\ Hu for useful discussions. This work was supported by the U.\! S.\ DOE,
Office of Science, under contract \# DE-AC02-06CH11357.


\begin{thebibliography}{99}
\bibitem{footnote1122}This alternating solution with the periodicity 1-2-1-2-$\ldots$ is
not the only possible state.  The solution with the periodicity
1-1-2-2-1-1-$\ldots$ also can be built in a similar way.
\end{thebibliography}
\end{document}